\documentclass[twocolumn,aps,prb,amsmath,amssymb]{revtex4}
\usepackage{graphicx}% Include figure files
\begin{document}

\title{In defence of negative temperature}

\author{J. Poulter}
\email{julian@swu.ac.th}
\affiliation{Department of Physics, Faculty of Science,
Srinakarinwirot University, 114 Sukumvit 23, Bangkok 10110, Thailand}

\date{\today}

\begin{abstract}
This pedagogical comment highlights three misconceptions concerning the usefulness of the concept of negative temperature; being derived from the 
usual, often termed Boltzmann, definition of entropy. 
First, both the Boltzmann and Gibbs entropies must obey the same thermodynamic consistency relation. 
Second, the Boltzmann entropy does obey the second law of thermodynamics. 
Third, there exists an integrating factor of the heat differential with both definitions of entropy.  

\end{abstract}

\maketitle

\section{Introduction}

The concept of negative temperature first arose, for spin systems, in the $1950$s \cite{PP51, Ramsey56} and has since been useful for understanding
a variety of systems with spontaneous magnetic order \cite{Hakonen94}. The essential feature of these systems is an energy spectrum that is
bounded above, as well as below. This means that the canonical partition function can undergo analytic continuation to negative temperatures. 
The total energy of the system at negative temperature is higher than that at infinite temperature.

More recently, negative temperatures have been observed relating to motional degrees of freedom \cite{Braun13}. An ensemble of particles with
an attractive interaction, or negative pressure, was prepared confined to a band with upper and lower bounds on kinetic energy. Since the system 
was observed to be stable, not imploding, we can conclude that its temperature must be negative.

Since the idea of temperatures being negative is unintuitive, there has arisen a controversy as to whether it may be advantageous to use an alternative
definition of entropy that guarantees only positive temperatures. In ref. \onlinecite{Dunkel14} it has been suggested that we should use the Gibbs, or 
volume, entropy since it increases monotonically with energy. 

To elaborate we consider an isolated system with energy $E$ in the microcanonical
ensemble \cite{Huang}. We denote as $\Gamma(E)$ the number of microstates that can be accessed by the system. It is important to recognise 
that $\Gamma(E)$ is a distribution. Especially if the energy spectrum is continuous then it gives the number of microstates in a distribution bin of some width
$\Delta$. The cumulative distribution $\Omega(E)$ represents the sum of $\Gamma(E)$ from the ground state up to energy $E$; we assume that the
energy spectrum is bounded below.

The two definitions of entropy are the Boltzmann, or surface, entropy
\begin{equation}
S_B = k_B \ln \Gamma(E)
\label{e:SB}
\end{equation} 
and the Gibbs, or volume, entropy
\begin{equation}
S_G = k_B \ln \Omega(E)
\label{e:SG}
\end{equation}
where $k_B$ is Boltzmann's constant. The usual situation in statistical mechanics is that the energy spectrum is not bounded above and the
value of $\Gamma(E)$ increases exponentially. In these circumstances, both entropy definitions are expected to provide the same results \cite{Huang}.

The temperature $T$ of the isolated system is given by
\begin{equation}
\frac{1}{T} = \frac{\partial S}{\partial E}
\label{e:temperature}
\end{equation}
Since the cumulative distribution $\Omega(E)$ is uniformly increasing, the Gibbs temperature is necessarily positive; $T_G > 0$. In contrast, for systems
with an energy spectrum bound above, it is the case that $\Gamma(E)$ is a decreasing function for high energy and this means that the 
Boltzmann temperature $T_B$ can be negative.

The main proposal of ref. \onlinecite{Dunkel14} is that it is always appropriate to use the Gibbs entropy and reject the Boltzmann entropy. This has
created a controversy with support for the proposal \cite{Hilbert14, DH14, Campisi15, Hanggi15} as well as criticism in support of keeping 
the Boltzmann entropy and negative temperatures \cite{Vilar14, Frenkel14, Schneider14, Swendsen14,SwendsenPRE,Wang15,Swendsen15,Buonsante}. In ref. \onlinecite{Vilar14}, for instance, it is shown that using the Gibbs entropy can mean heat flowing from cold to hot as well as the temperature $T_G$ not being intensive. 

The aim of this short note is not to review the controversy in detail but just to point out three simple observations in support of negative temperatures
and thus provide a clean bill of health for the Boltzmann entropy.
First, it is not true that thermodynamic consistency is ever violated if we use $S_B$. Second, use of $S_B$ does not violate the second law of thermodynamics. Third, we can find an integrating factor for the heat differential using either definition of entropy. 
 
\section{Thermodynamic consistency}

Suppose we have three variables $S$, $E$ and $h$ and that some function of them obeys $f(S,E,h)=0$. Then
\begin{equation}
df = f_S dS + f_E dE + f_h dh = 0
\end{equation}
where the coefficients are partial derivatives. First, fixing $h$, we can prove that
\begin{equation}
\left( \frac{\partial S}{\partial E} \right)_h \left( \frac{\partial E}{\partial S} \right)_h = 1
\label{e:relation1}
\end{equation}
Then, fixing each variable in turn, we derive three equations from which we can eliminate the derivatives of $f$ and arrive at
\begin{equation}
\left( \frac{\partial S}{\partial E} \right)_h
\left( \frac{\partial E}{\partial h} \right)_S
\left( \frac{\partial h}{\partial S} \right)_E = -1
\label{e:relation2}
\end{equation}
These equations are very well known from thermodynamics \cite{Huang}. With $T$ defined by
\begin{equation}
\frac{1}{T} = \left( \frac{\partial S}{\partial E} \right)_h
\label{e:T}
\end{equation}
we can see that, from (\ref{e:relation2}),
\begin{equation}
\frac{1}{T} \left( \frac{\partial E}{\partial h} \right)_S \left( \frac{\partial h}{\partial S} \right)_E = -1
\end{equation}
and then, using a variant of (\ref{e:relation1}),
\begin{equation}
T \left( \frac{\partial S}{\partial h} \right)_E = - \left( \frac{\partial E}{\partial h} \right)_S
\label{e:consistency}
\end{equation}

So far this is just mathematics; $S(E,h)$ can be any single-valued section of a differentiable function without singular points. 
We continue by considering an isolated system with entropy $S$, 
fixed energy $E$ and subject to an external influence $h$. Thermodynamic consistency says that \cite{Dunkel14}
\begin{equation}
T \left( \frac{\partial S}{\partial h} \right)_E = - \left( \frac{\partial E}{\partial h} \right)_S = - \left\langle \frac{\partial H}{\partial h} \right\rangle
\label{e:consistency}
\end{equation}
where the $H$ is the Hamiltonian and the average is evaluated in the microcanonical ensemble. The influence $h$ can of course be interpreted 
as a vector if there are many components.

First, let us consider a classical system with Hamiltonian $H(p,q)$. As usual, $p$ and $q$ are vectors for the canonical momenta and coordinates.
In three space dimensions with $N$ particles we have vectors in a $3N$-dimensional space. The energy spectrum is not discrete and we need to
identify $\Gamma(E)$ as the number of accessible microstates in a distribution bin of some width $\Delta$. We can write \cite{Huang}
\begin{equation}
\Gamma(E) = \int_{E<H<E+\Delta} dpdq
\end{equation} 
In order to make this dimensionless it is usual to divide by Planck's constant $3N$ times but this detail will not prove important.

The ensemble average of the Hamiltonian is
\begin{align}
\langle H \rangle &= \frac{1}{\Gamma(E)} \int_{E < H < E+\Delta} H dp dq\\
&= E + \frac{1}{\Gamma(E)} \int_{E<H<E+\Delta} (H-E) dp dq
\end{align}
Then we can use the bound 
\begin{equation}
\left| \frac{1}{\Gamma(E)} \int_{E<H<E+\Delta} (H-E) dpdq \right| < \Delta 
\end{equation}
to deduce that
\begin{equation}
E < \langle H \rangle < E+\Delta
\end{equation}
It is important that all that has physical significance is independent of the bin width $\Delta$. Although we choose $\Delta$ to be very small it
should not depend on the number of particles $N$. For example $\Delta = e^{-N}$ would not provide us with a distribution. We must insist that the 
total number of distribution bins $n$ is very small in relation to the number of particles; that is $n \ll N$. Now, since the energy $E$ is an extensive quantity, we have that $E \sim N$ and can treat $\Delta \ll E $ in the thermodynamic limit \cite{Huang}. Thus
\begin{equation}
E = \langle H \rangle
\end{equation}
with sufficiently large $N$.

Returning to consider Eq. (\ref{e:consistency}) we have that
\begin{equation}
\left( \frac{\partial E}{\partial h} \right)_S = \left( \frac{\partial}{\partial h} \frac{1}{\Gamma(E)} \int_{E<H<E+\Delta} H dp dq \right)_S
\label{e:fixS}
\end{equation}
We can imagine a second microcanonical ensemble with the same entropy $S$ but subject to an influence $h+dh$. The energy of this second ensemble
must be $E+dE$ with
\begin{equation}
dE = \left( \frac{\partial E}{\partial h} \right)_S dh
\end{equation}
The way to fix the entropy in Eq. (\ref{e:fixS}) is to simply integrate over the microstates accessible to the first unperturbed ensemble.
Then the derivative only acts on the integrand and
\begin{equation}
\left( \frac{\partial E}{\partial h} \right )_S = \frac{1}{\Gamma(E)} \int_{E<H<E+\Delta} \frac{\partial H}{\partial h} dpdq
\end{equation}
and this is precisely what we desire. Of course, fixing the set of accessible microstates also fixes $\Gamma(E)$ and the entropy $S_B = k_B \ln \Gamma(E)$.
Furthermore we have also fixed the cumulative distribution $\Omega(E)$ and the entropy $S_G=k_B \ln \Omega(E)$. 
It does not matter which definition of entropy we use to prove thermodynamic consistency. 

It is worthwhile to emphasize here that we are just using the obvious statistical expression
\begin{equation}
\frac{\partial}{\partial h} \langle H \rangle = \left \langle \frac{\partial H}{\partial h} \right \rangle
\end{equation}
where the averages over the microstates must have exactly the same meaning. The mapping from  a set of microstates to some thermodynamic entropy
function is essentially free, although perhaps unphysical in general.

For systems with a discrete energy spectrum, it is most convenient to put one energy level into one distribution bin, although we might in principle
put more than one. In any case we just need to sum over the levels in the bin. For spin systems we also need to replace the integrals over $p$ and $q$
with discrete sums. However, regardless of these details, we fix the entropy by a trace over the microstates accessible to the unperturbed system.
We have thermodynamic consistency for all systems using either $S_B$ or $S_G$.

This can be illustrated by the example of a gas of Ising spins.
The Hamiltonian is
\begin{equation}
H = -h \sum_{i=1}^{N} s_i
\end{equation}
Here $h$ is an external field with units of energy. The Ising spins take values $s_i = \pm 1$. This is basically the two-level system commonly introduced
to illustrate the idea of negative temperature \cite{Dunkel14,Campisi15,Vilar14,Frenkel14,Swendsen14,Wang15,Swendsen15}. The energy spectrum is discrete; with $N_{\uparrow}$ up spins with $s_i=+1$ and $N_{\downarrow}$ down spins
\begin{equation}
E =\langle H \rangle = -h(N_{\uparrow}-N_{\downarrow})
\end{equation}
Since the spectrum is discrete we  put one energy level into one distribution bin and perform discrete sums over the spins. 
With $N_{\uparrow}$ up spins the number of accessible microstates with energy $E$ is
\begin{equation}
\Gamma(E) = \frac{N!}{N_{\uparrow}! N_{\downarrow}!}
\end{equation}
where $N_{\uparrow}=\frac{1}{2} (N - \frac{E}{h})$ and $N=N_{\uparrow}+N_{\downarrow}$. The cumulative distribution is
\begin{equation}
\Omega(E) = \sum_{n=0}^{N_{\downarrow}} \frac{N!}{n! (N-n)!}
\end{equation}
As we expect, fixing the value of $N_{\uparrow}$ fixes $\Gamma(E)$, $S_B$, $\Omega(E)$ and $S_G$. Our thermodynamic consistency equation
reads
\begin{equation}
T \left( \frac{\partial S}{\partial h} \right)_E = - \left( \frac{\partial E}{\partial h} \right)_{N_{\uparrow}} = N_{\uparrow}-N_{\downarrow}
\end{equation}
where we can use either definition of entropy.

Our general conclusion is that
\begin{equation}
T_B \left( \frac{\partial S_B}{\partial h} \right)_E = T_G \left( \frac{\partial S_G}{\partial h} \right)_E
\label{e:either}
\end{equation}
in contradiction to a statement given in ref. \onlinecite {Dunkel14}. A proof that the Boltzmann entropy $S_B$ does obey thermodynamic 
consistency has also been given in ref. \onlinecite{Frenkel14}. The technique included the saddle-point method which is certainly valid where
$N$ is sufficiently large. For the gas of Ising spins the energy, or magnetic enthalpy, is simply
\begin{equation}
E = -Nh \tanh \beta h
\end{equation}
where $\beta = \frac{1}{k_B T}$ and the temperature $T$ is interpreted as the Boltzmann temperature $T_B$. If more than half of the spins
are down then the energy is positive and the temperature must be negative. This result can be derived simply from either the canonical or microcanonical
ensemble.

\section{Second law of thermodynamics}

We aim to prove here that the Boltzmann entropy $S_B$ is always consistent with the second law of thermodynamics. In ref. \onlinecite{Hilbert14} it is
suggested that this may not be the case under certain conditions. Following ref. \onlinecite{Hilbert14} we state the second law, according to Planck, as
\begin{equation}
S(E_1 + E_2) \ge S_1(E_1) + S_2(E_2)
\label{e:secondlaw}
\end{equation}
The terms on the right are the entropies of two microcanonical ensembles with respective energies $E_1$ and $E_2$. We imagine coupling these systems and waiting long enough for a new thermodynamic equilibrium. Then the term on the left is the entropy of the composite system which has energy
$E_1+E_2$. The equality occurs if the uncoupled systems have the same temperature. Otherwise the entropy increases.

The uncoupled systems have distributions $\Gamma_1(E_1)$ and $\Gamma_2(E_2)$ for their numbers of  accessible microstates. Thus, before coupling, the
number of microstates accessible to the composite system is $\Gamma_1(E_1)\Gamma_2(E_2)$. After coupling, the distribution for the composite system will
be \cite{Huang}
\begin{equation}
\Gamma(E_1+E_2) = \sum_{i=1}^n \Gamma_1(E_i) \Gamma_2(E_1+E_2-E_i) 
\label{e:coupled}
\end{equation}
with a sum over $n \ll N$ distribution bins. The uncoupled systems have numbers of particles proportional to $N$ and their distributions use the same
numbers of bins. On the left we should best use a bin width $2\Delta$ double that on the right. With a discrete spectrum we can take a simple sum over the 
energy levels. Otherwise, as long as we maintain $n \ll N$, we can take $n$ to be very large and write
\begin{equation}
\Gamma(E_1+E_2) = \int_{E_{01}}^{E_1+E_2-E_{02}} \Gamma_1(E) \Gamma_2(E_1+E_2-E) dE
\end{equation}
where $E_{01}$ and $E_{02}$ are the ground state energies of the uncoupled systems.

In all cases we have that
\begin{equation}
\Gamma(E_1+E_2) \ge \Gamma_1(E_1) \Gamma_2(E_2)
\label{e:distributions}
\end{equation}
since all the terms in the sum are nonnegative. If we now define entropy according to $S=k_B \ln \Gamma$ then we arrive at the
inequality (\ref{e:secondlaw}). This is a proof that using the Boltzmann entropy never violates the second law of thermodynamics.
In fact, the sum (\ref{e:coupled}) is dominated by one particular term \cite{Huang}
\begin{equation}
\Gamma(E_1+E_2) \simeq \Gamma_1(\overline{E}_1) \Gamma_2(\overline{E}_2)
\end{equation}
where $\overline{E}_1$ and $\overline{E}_2$ are the energies corresponding to thermal equilibrium.
This leads to 
\begin{equation}
S(E_1+E_2) = S_1(\overline{E}_1) + S_2(\overline{E}_2)
\label{e:additivity}
\end{equation}
in the thermodynamic limit; in agreement with the postulate of the additivity of entropy \cite{Swendsen14,SwendsenPRE} which is essential with regard to prediction of the equilibrium energies.

The density of states $\omega(E)$ is defined \cite{Huang} by
\begin{equation}
\Gamma(E) = \omega(E) \Delta
\end{equation}
Provided the distribution bin width $\Delta$ is sufficiently small, we can write $\omega(E) = \frac{\partial \Omega(E)}{\partial E}$. In ref. \onlinecite{Hilbert14} it is written that $\Gamma=\epsilon \omega$ where $\epsilon$ is described as a parameter that ensures that $\Gamma$ is 
dimensionless. In fact, $\epsilon$ must be interpreted as a distribution bin width; not freely. In terms of the density of states, inequality
(\ref{e:distributions}) reads
\begin{equation}
\omega(E_1+E_2) \ge  \omega_1(E_1) \omega_2(E_2) \frac{\Delta_1 \Delta_2}{\Delta}
\end{equation}
We can define \cite{Huang} entropy via $S=k_B \ln \omega$ and, with say $\Delta_1=\Delta_2=\frac{1}{2}\Delta$,
\begin{equation}
S(E_1+E_2) \ge S_1(E_1) + S_2(E_2) + k_B \ln \frac{1}{4}\Delta
\end{equation}
The third term must be negligible in the thermodynamic limit since entropy is extensive, $S \sim N$. As remarked above, choosing say
$\Delta = e^{-N}$ is not possible since we would not have a distribution for $\Gamma(E)$. The value of $\Delta$ must not depend on $N$ and
both $S=k_B \ln \Gamma$ and $S= k_B \ln \omega$ obey the second law.

Much discussion as to whether the Gibbs entropy $k_B \ln \Omega$ obeys the second law can be found elsewhere \cite{Hilbert14,Vilar14,Frenkel14,Swendsen14,SwendsenPRE,Wang15}. In ref. \onlinecite{Hilbert14} it is proven that
\begin{equation}
\Omega(E_1+E_2) \ge \Omega_1(E_1) \Omega_2(E_2)
\end{equation}
and we can conclude that the Gibbs entropy $S_G$ does satisfy the second law as stated in Eq. (\ref{e:secondlaw}).
Nevertheless, the crucial issue concerns the additivity of entropy as expressed in Eq. (\ref{e:additivity}).  In the event that the Boltzmann entropy
in increasing with energy it is well known \cite{Ma} that we can prove that $S_B$ and $S_G$ provide the same thermodynamic predictions. 
It is in the case that $S_B$ is not increasing that difficulties arise \cite{Swendsen14,SwendsenPRE,Swendsen15}.

Following the construction in ref. \onlinecite{Hilbert14} we can write exactly
\begin{equation}
\begin{aligned}
\Omega(\overline{E}_1&+\overline{E}_2) = \Omega_1(\overline{E}_1) \Omega_2(\overline{E}_2)\\
&+\int_{0}^{\overline{E}_2-E_{02}} \Gamma_1(E+\overline{E}_1)  \int_{E_{02}}^{\overline{E}_2-E} \Gamma_2(E')dE'dE\\
&+\int_{0}^{\overline{E}_1-E_{01}}\Gamma_2(E+\overline{E}_2)\int_{E_{01}}^{\overline{E}_1-E}\Gamma_1(E')dE'dE
\end{aligned}
\label{e:everything} 
\end{equation}
The Gibbs entropy will only give a correct thermodynamic prediction if the second and third lines do not grow exponentially with system size.
Looking at the second line above, we can image that $\Gamma_2(E')$ is not monotonic and possesses a peak at some value $E'=\overline{E}'$.
If $\overline{E}' < \overline{E}_2$ then the line will indeed provide a contribution that grows exponentially with size. This destroys the utility
of the Gibbs entropy. 

To illustrate we can use the gas of Ising spins. It is convenient here to parameterise the energy with the number $N_\downarrow$ of down spins;
the ground state has $N_\downarrow=0$. We couple two systems with $N_1$ and $N_2$ spins where $N=N_1+N_2$ and $N_1\sim N_2 \sim N$.
The total number of down spins $N_\downarrow=N_{\downarrow 1}+N_{\downarrow 2}$ is conserved. In thermal equilibrium we have
\begin{equation}
\frac{\overline{N}_{\downarrow 1}}{N_1} = \frac{\overline{N}_{\downarrow 2}}{N_2} = \frac{N_\downarrow}{N}
\end{equation}
Translating the second line of (\ref{e:everything}), we have
\begin{equation}
\sum_{m_1=1}^{\overline{N}_{\downarrow 2}} \Gamma_1(m_1+\overline{N}_{\downarrow 1})
\sum_{m_2=0}^{\overline{N}_{\downarrow 2}-m_1} \Gamma_2(m_2)
\label{e:translation}
\end{equation}
where
\begin{equation}
\Gamma_2(m_2) = \frac{N_2 !}{m_2 ! (N_2 - m_2)!} \simeq e^{N_2 f(x)}
\end{equation}
with $x=\frac{m_2}{N_2}$ and $f(x)=-x \ln x - (1-x)\ln(1-x)$. It is a simple matter to show that $f(x)$ has a maximum at $x=\frac{1}{2}$,
or $m_2 = \frac{1}{2} N_2$. This will contribute something large to the sum (\ref{e:translation}) if $\overline{N}_{\downarrow 2} > \frac{1}{2} N_2$
or, equivalently, $N_\downarrow > \frac{1}{2} N$ which is precisely where the temperature  is negative and the Gibbs entropy fails to provide 
any thermodynamic prediction of the equilibrium state. A numerical evaluation of this has been given in ref. \onlinecite{Swendsen15} where it is
also clear that the Gibbs temperature $T_G$ is not intensive, depending on system size contrary to common sense.

\section{Integrating factor for the heat differential}

In thermodynamics the heat differential $dQ$ is not exact \cite{Huang}. The integrating factor $\frac{1}{T}$, where $T$ is temperature, allows us to use an exact differential 
\begin{equation}
dS = \frac{dQ}{T}
\end{equation}
where $S$ is entropy. In ref. \onlinecite{Campisi15} it is suggested that this might not be valid in every case unless we use the Gibbs entropy $S_G$
and the Gibbs temperature $T_G$.

The work done, on a system, by an external influence $h$ is
\begin{equation}
dW = \left\langle \frac{\partial H}{\partial h} \right\rangle dh
\end{equation}
where we follow the notation above. Thus
\begin{equation}
dQ = dE - dW = dE + T \left( \frac{\partial S}{\partial h} \right)_E dh
\label{e:firstlaw}
\end{equation}
using Eq. (\ref{e:consistency}). We know that we can use either the Boltzmann or Gibbs definitions of entropy, and temperature, here.

Suppose we use $S=k_B \ln \Omega$; the Gibbs definition. Then, since $T=\left( \frac{\partial E}{\partial S} \right)_h$,
\begin{equation}
dQ = dE+\frac{1}{\omega} \frac{\partial \Omega}{\partial h}dh
\end{equation}
where $\omega=\frac{\partial \Omega}{\partial E}$ is the density of states at fixed $h$ and the derivative of $\Omega(E,h)$ is at fixed energy.
We can see that $\omega$ is an integrating factor since
\begin{equation}
\omega dQ = \frac{\partial \Omega}{\partial E} dE + \frac{\partial \Omega}{\partial h} dh = d\Omega
\end{equation}
Continuing, ref. \onlinecite{Campisi15} correctly argues that  integrating factors for $dQ$ must be of the form $\frac{\partial}{\partial E} g(\Omega)$
where $g$ is differentiable. The law of ideal gases then identifies a unique solution $g(\Omega)=k_B \ln \Omega$.

The point now is that we can return to Eq. (\ref{e:firstlaw}) and use the definition $S=k_B \ln \Gamma$ for the entropy and simply replace $\Omega$
with $\Gamma$ in all expressions. We can do this since both definitions obey the same thermodynamic consistency relations; we refer to Eq. (\ref{e:either}). In summary, both entropy definitions provide an integrating factor $\frac{1}{T}=\frac{\partial S}{\partial E}$.

As a matter of fact we can replace $\Omega$ with any $\Xi$ as long as, in the case of the ideal gas,
\begin{equation}
\Xi \sim E^p
\end{equation}
with $p=\frac{3N}{2}-c$ where $c$ is constant, or unimportant in the thermodynamic limit. We have $c=0$ with $\Omega$ and $c=1$ with $\Gamma$.

\section{Discussion}

We have proven three important properties of the Boltzmann entropy that have been challenged in the literature. It always obeys thermodynamic consistency, the second law of thermodynamics and provides an integrating factor for the heat differential.

One important theme is that the number of accessible microstates $\Gamma(E)$ should be regarded as a distribution. It does not necessarily have to 
be differential or even continuous although we usually assume that it is.

A possible further comment is that we have the microcanonical ensembles subject to an external influence when they are supposed to be
isolated. Also, for the gas of Ising spins, we have no interactions between the particles and may wonder how two systems can mix or couple and find 
a new thermal equilibrium. The usual answer to this is that the particles have weak interactions.

These points might be addressed by looking at the example of an Ising model in one dimension with Hamiltonian
\begin{equation}
H = - J \sum_{i=1}^{N-1} s_i s_{i+1}
\end{equation}
which has also proven useful in ref. \onlinecite{Wang15}.
The exchange interaction $J$ is positive, as large as we like, and there is no external influence. To solve this problem we can imagine a bond with parallel spins as represented by another Ising variable $\sigma =+1$. A bond with spins not parallel has $\sigma=-1$. We take $N$ to be very large and do not
worry about boundary effects.

The degeneracy of the energy level $E=-J(N_{+}-N_{-})$ is given by
\begin{equation}
\Gamma(E) = 2 \frac{N!}{N_{+}!N_{-}!}
\end{equation}
The factor of $2$ is for global inversion of the spins. $N_{+}$ is the number of bonds with parallel spins and $N_{-}=N-N_{+}$.
In the same way as for the gas of Ising spins we can solve this model by just writing $J$ instead of $h$. We have that the total energy is
\begin{equation}
E = - N J \tanh \beta J
\end{equation}
If more than one half of the bonds have spins not parallel then this energy is positive and the temperature is negative, provided we use the usual Boltzmann definition. 

We can imagine some divine intervention that arranges a majority of bonds with spins not parallel and then isolates the system very quickly. 
The spins would then come to thermal equilibrium, through their interactions, at a negative temperature. 

A similar illustration can easily be considered in higher dimensions. For instance, a square lattice Ising model, with all exchange interactions $J$ positive,
can be isolated at negative temperature if a majority of bonds have spins not parallel. A perfect antiferromagnetic arrangement would indicate
the extreme case of temperature negative zero.

An important utility of the concept of negative temperature is that we can use the canonical ensemble. Where the energy spectrum is bounded above, 
the partition function \cite{Huang} can be defined at negative temperature by simple analytical continuation. Since the microcanonical
ensemble is generally much less mathematically tractable, this is a powerful point.

The Gibbs temperature is guaranteed to be positive and this is intuitively attractive. Nevertheless, we cannot use the canonical ensemble and there are serious issues about thermodynamic prediction in the case that $\Gamma(E)$ is not monotonically increasing .

\end{document}